\documentclass[aapm,graphicx,mph,reprint]{revtex4-1}
\pdfoutput=1
\usepackage{amsmath,amsfonts,amssymb}
\usepackage{graphicx}
\usepackage{setspace}
\usepackage{tocloft}

\usepackage{lineno}
\usepackage{siunitx}

\DeclareMathOperator*{\argmin}{argmin\,}

\begin{document} 
\title{Imaging of Fiber-Like Structures in Digital Breast Tomosynthesis}


\author{Sean D. Rose}
\affiliation{University of Wisconsin, Dept. of Medical Physics, 1111 Highland Avenue, Madison, WI, 53705}
\author{Emil Y. Sidky}
\author{Ingrid Reiser}
\author{Xiaochuan Pan}
\affiliation{University of Chicago, Dept. of Radiology MC-2026, 5841 S. Maryland Avenue, Chicago IL, 60637}


\keywords{Digital Breast Tomosynthesis, Breast Imaging, Image Reconstruction, Fiber Signals}
\begin{abstract}
Fiber-like features are an important aspect of breast imaging. Vessels and ducts
are present in all breast images, and spiculations radiating from a mass can
indicate malignancy. Accordingly, fiber objects are one of the three types
of signals used in the American College of Radiology digital mammography (ACR-DM) accreditation phantom. 
This work focuses on the image properties of fiber-like structures in digital breast
tomosynthesis (DBT) and how image reconstruction can affect their appearance.
The impact of DBT image reconstruction
algorithm and regularization strength on the conspicuity of fiber-like signals
of various orientations is investigated in simulation. 
A metric is developed to characterize this orientation dependence and allow for quantitative comparison
of algorithms and associated parameters in the context of imaging fiber signals.
The imaging properties of fibers, characterized in simulation, are then
demonstrated in detail with physical DBT data of the ACR-DM phantom.
The characterization of imaging of fiber signals is
used to explain features of an actual clinical DBT case.
For the algorithms investigated, at low regularization
setting, the results show a striking variation in conspicuity as a function of orientation in the viewing plane. In particular, the conspicuity of fibers nearly aligned with the
plane of the X-ray source trajectory is decreased relative to more obliquely oriented fibers. Increasing regularization strength mitigates this orientation
dependence at the cost of increasing depth blur of these structures. 
\end{abstract}

\maketitle




\section{Introduction}
\label{sec:intro}  
Over the past decade, digital breast tomosynthesis (DBT) has seen widespread
clinical adoption as either a supplement to, or replacement for, full-field digital
mammography (FFDM) for breast cancer screening. Similar to FFDM, the generic DBT configuration
employs a static detector and breast compression. Unlike FFDM, the X-ray source sweeps a
15$^\circ$ to 50$^\circ$ arc and multiple projections are acquired.
The additional projections provide data for some resolution of 3D structures that may be
superposed in a 2D mammography image.

The formation of the volume is achieved by tomographic image reconstruction algorithms.
Both direct and iterative algorithms\cite{Sechopoulos2013} have been adapted to DBT,
but due to the limited-angle scanning arc, the volume resolution properties are highly
anisotropic. ``In-plane'' slice resolution is the same as that of FFDM
while ``depth'' resolution is low and object-dependent, where ``in-plane'' refers to
volume slices parallel to the detector and ``depth'' refers to the direction perpendicular
to the detector.

A substantial body of work has been devoted to the characterization and optimization
of DBT acquisition \cite{Zhou2007,Hu2008,Zhao2009,Chawla2009,Sechopoulos2009,Saunders2009,Park2010,Reiser2010,Richard2010,Gang2011,Lu2011,Young2013,Tucker2013,Chan2014,Goodsitt2014} and image reconstruction
algorithm\cite{Wu2004,Mertelmeier2006,Zhou2007,Zhao2009,VandeSompel2011,Zeng2015,Sanchez2015a,Rose2017} parameters.
The majority of the work has focused on two specific clinically relevant
tasks: 
the detection and visual assessment of small, high-contrast microcalcifications and
of large, low contrast lesions
modeled using rotationally symmetric signals of varying size and contrast.
Such signals, however, do not address in-plane anisotropy of DBT.

In-plane anisotropy in DBT imaging properties results from the direction of X-ray
source travel (DXST). This in-plane anisotropy is conspicuous in the imaging of
micro-calcifications and fiber-like signals. Micro-calcifications can display strong
undershoot artifacts and ghosting in adjacent slices, and fiber-like
signals enhance better for orientations oblique to the DXST than they do when
aligned parallel to the DXST.\cite{MaidmentRSNA2017}
While the image of a micro-calcification in a DBT slice appears anisotropic,
its image properties vary little with orientation of the microcalcification itself due to
its compactness. 
Fiber-like signals, however, have strong image property dependences on signal orientation
due to their shape and low contrast. Detection of such signals is significant because it impacts
visual assessment of spiculated lesions, Cooper's ligaments, and blood vessels.

In this work, we focus on characterization of DBT image reconstruction algorithms and associated parameters
based on the imaging properties of fiber-like signals. The characterization is performed in
simulation and a 
simulation-based image relative intensity metric is developed \cite{Rose2017}.
We demonstrate the use of the simulations and new metric 
to explain the imaging properties of fiber-like
structures in a physical phantom and DBT clinical case studies.
Section \ref{sec:methods} describes the DBT configuration, data model, image reconstruction
algorithms, and the relative intensity metric used to interpret the imaging of fiber-like signals;
Sec. \ref{sec:results} shows results for imaging of fiber-like signals in simulation,
physical phantom studies, and for a DBT clinical case; and
Sec. \ref{sec:conclusion} discusses the results and presents conclusions.

\section{Methods and materials}
\label{sec:methods}

\color{red}
\color{black}

\begin{figure}[t!]
	\centering
	\includegraphics[width=0.47\textwidth]{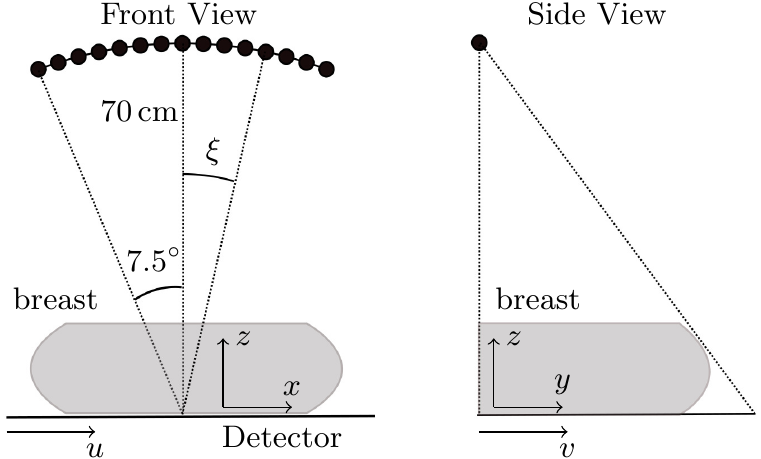}
\caption{\label{fig:geo} Illustration of the approximate fixed-detector DBT scan geometry.
The $x,y,z$ coordinates parametrize the reconstruction volume, and the $u,v$ coordinates
specify location on the detector. The X-ray source travels on a 15 degree circular arc,
centered on the middle of the long-axis ($u$ direction) of the detector and the edge of its short-axis
($v$ direction).
The radius of the scanning arc is $C=70$ cm. The X-ray source position is indicated by the
angle $\xi$. The long-axis of the detector measures at $L=28.7$ cm,
the short-axis is 23.3 cm, and the detector pixels are (0.14 mm)$^2$.}
\end{figure}

\subsection{DBT System Geometry}
\label{sec:config}
Data for both physical phantom and patient studies was acquired with
a Hologic Selenia Dimensions DBT scanner,
in which 15 projections are obtained over a scanning arc which is approximately 15 degrees.
With the actual scanner, both X-ray source and flat-panel detector move with each projection. The
geometric configuration data are recorded with each scan and are stored in the form of projection matrices.
These matrices map 3D spatial coordinates of the volume to 2D detector coordinates for
each projection view \cite{Li2010}. 

To explain the DBT data model and image reconstruction algorithms,
we employ an approximate simplified geometric model for the DBT scan shown in
Fig. \ref{fig:geo}. In this model, the detector is fixed;
the X-ray source moves on a circular arc of 15 degrees, acquiring 15 projections in 1 degree increments;
the center of the scanning arc lies on the detector; and the detector pixel size is 0.14 mm $\times$ 0.14 mm.
\color{red} The various geometric parameters of the scan are also shown in the schematic
in Fig. \ref{fig:geo}. \color{black}

While the data model and image reconstruction theory is explained with the approximate
stationary-detector model, the image reconstruction results are obtained with the
precise scan geometry provided by the projection matrices.

\subsection{DBT Data model}
\label{sec:datamodel}
The X-ray transmission data are modeled with the Beer-Lambert law
\begin{equation}
\label{Beer}
I(\xi,u,v) = I_0 \exp \left( -g(\xi,u,v) \right),
\end{equation}
where $I_0$ is the incident X-ray fluence; $I(\xi,u,v)$ is the transmitted fluence for
X-ray source angle $\xi$ and detector pixel location $(u,v)$. The function $g(\xi,u,v)$
represents the line integrals over the object function $f(\vec{r})$
\begin{equation}
\label{Xraymodel}
g(\xi,u,v) =  \int_{0}^{\infty} f \left( \vec{r}_0(\xi)+ t \hat{\theta}(\xi,u,v) \right) \mathrm{d}t
\end{equation}
where $f(\vec{r})$ represents an energy-averaged map of the linear X-ray attenuation coefficient;
the X-ray source position is $\vec{r}_0(\xi)$; and the unit vector $\hat{\theta}(\xi,u,v)$ indicates
the direction of a ray originating from $\vec{r}_0(\xi)$ and incident on the detector at $(u,v)$.
Using the coordinate system shown in Fig. \ref{fig:geo}, the source position can be parameterized by
\begin{align*}
	\vec{r}_0( \xi) =  (C \sin \xi, 0, C \cos \xi)
\end{align*}
and the detector pixel locations by 
\begin{align*}
	\vec{d}(u,v) =  (u - L/2, v, 0)
\end{align*}
where $C=70$ cm, the radius of the source trajectory, and $L=28.7$ cm, the length of the detector in the $x$ direction.
The unit vector $\hat{\theta}(\xi,u,v)$ can then be written
\begin{align*}
	\hat{\theta}(\xi,u,v) = \frac{\vec{d}(u,v) - \vec{r}_0(\xi)}{|\vec{d}(u,v) - \vec{r}_0(\xi)|}.
\end{align*}
The $u$ coordinates are shifted by $L/2$ with respect to the volume coordinates because
the origin of the detector is taken to be at a corner, while the origin of the volume Cartesian
coordinates coincides with the middle of the detector edge adjacent to the chest wall.

The noisy simulation data, which is generated for the results in Sec. \ref{sec:results},
use the Beer-Lambert model in Eq. (\ref{Beer}) as a mean value for a Poisson distribution,
where the noise level is specified by selecting the total number of photons, $I_0$, incident
at each detector bin. No correlation between detector bins is assumed.

For image reconstruction, the DBT transmission data are processed by inverting Eq. (\ref{Beer})
\begin{equation*}
g(\xi,u,v) = - \log \left( \frac{I(\xi,u,v)}{I_0} \right),
\end{equation*}
to obtain the projection data.
Image reconstruction algorithms are based on the mathematical inversion of the X-ray projection
model in Eq. (\ref{Xraymodel}).

\subsection{DBT image reconstruction algorithms}
\label{sec:recon}


For the results presented in Sec. \ref{sec:results}, four image reconstruction algorithms are
compared. The first two are of the form of filtered back-projection (FBP). For background
on application of FBP in CT, the reader is referred to Ref. \onlinecite{Kak1988}, and for implementation of FBP specific to DBT,
Ref. \onlinecite{Mertelmeier2017}.
We also consider two common forms of penalized, least-squares optimization that we have
already investigated in the context of signal detection in DBT \cite{Sanchez2015a,Rose2017}.

For the FBP implementations, the
general form of filtering considered involves only filtering along the detector $u$-direction with
the one-dimensional Fourier transform (FT) and inverse Fourier transform (IFT)
\begin{equation}
\label{filter}
h(\xi,u,v) = \int 
e^{2 \pi i u \nu} H(\nu) \int e^{-2 \pi i \nu u^\prime} g(\xi,u^\prime,v) d u^\prime d\nu,
\end{equation}
where integration over $u^\prime$ performs the FT; $H(\nu)$ is the filter function;
integration over $\nu$ performs the IFT; and $h(\xi,u,v)$ is the filtered projection data.

The reconstructed volume is obtained from $h(\xi,u,v)$ by back-projection,
\begin{equation}
\label{backproj}
f_\text{recon}(\vec{r}) = \int^{\xi_\text{max}}_{\xi_\text{min}} h\left[ \xi, u(\xi,\vec{r}),v(\xi,\vec{r})\right] d\xi,
\end{equation}
where $u(\xi,\vec{r})$ and $v(\xi,\vec{r})$ are the coordinates on the detector pointed to by a ray
originating at source position $\vec{r}_0(\xi)$ and passing through $\vec{r}=(x,y,z)$, i.e.
\begin{align*}
u[\xi,(x,y,z)] & = C\sin\xi + (x-C\sin\xi)\frac{C\cos\xi}{C\cos\xi-z} +\frac{L}{2}, \\
v[\xi,(x,y,z)] & = y\frac{C\cos\xi}{C\cos\xi-z}
\end{align*}

\textbf{FBP} - The filter for the first FBP algorithm is a ramp filter with a Hanning
apodizing window
\begin{align*}
H_\text{FBP} &= |\nu| H_\text{Hanning}(\nu),\\
H_\text{Hanning}(\nu) &= \begin{cases} \frac{1}{2}\left(1 +
\cos \left(\frac{\pi \nu }{\nu_H} \right)\right) & |\nu| \le \nu_H \\
0 &  |\nu| > \nu_H
\end{cases},
\end{align*}
where $\nu_{H}$ is the cutoff frequency of the filter.
The cutoff frequency $\nu_{H}$ is specified as a fraction of the detector's Nyquist frequency $\nu_{max}$
\begin{align*}
	&\nu_{max} = \frac{1}{2 \Delta u} \\
	&\nu_c = \frac{\nu_H}{\nu_{max}}
\end{align*}
where $\Delta u$ is the detector bin size and $\nu_c$ is the fractional cutoff frequency.
For the studies in Sec. \ref{sec:results}, we investigate the impact
of $\nu_c$ on conspicuity of fiber-like signals.

\textbf{mFBP} - One strategy for designing analytic image reconstruction
for tomosynthesis involves combining back-projection BP with ramp-filtered FBP.
Such an admixture has been proposed for reduction of metal artifacts \cite{Gomi2009}.
A similar approach is to 
 modify the FBP filter in such a way that its value at low spatial frequency is boosted with
respect to the ramp \cite{Orman2006}.
For mFBP, short for modified FBP, we design a filter based on the latter idea. 
Our particular choice of filter function employs a shifted quadratic at low spatial frequency
\begin{align*}
&H_{\text{mFBP}} = H_{\delta}(\nu) H_{\text{Hanning}}(\nu)\\ 
&H_{\delta}(\nu) = \begin{cases} \delta \nu_\text{max}
( 1 + \nu^2/(2 \delta \nu_\text{max})^2 ) & |\nu | \leq 2 \delta \nu_\text{max} \\
			|\nu | & |\nu | > 2 \delta \nu_\text{max} \end{cases},
\end{align*}
where $\delta$ is specified in terms of a fraction of the detector's Nyquist frequency.
The filter $H_{\text{mFBP}}$ used for mFBP is the product of the boosted filter $H_\delta(\nu)$ and the Hanning
apodization window $H_\text{Hanning}(\nu)$ for $\nu_c = 1.0$.
For mFBP, $\delta$ is the only regularization parameter and is taken to be in the range $[0,\pi^{-1}]$.

At $\delta=0$, $H_\text{mFBP}(v)$ is identical to $H_\text{FBP}(v)$ for $\nu_c=1$.
The upper bound on the range $\delta = \pi^{-1}$ is determined by the condition
$\frac{ \mathrm{d}^2 H_\text{mFBP}}{\mathrm{d} \nu^2}(0) = 0$. This condition yields a flat
response at low frequencies (the first derivative at $\nu=0$ is zero because the filter function is symmetric).
At $\delta= \pi^{-1}$, $H_\text{mFBP}(v)$ becomes
a low-pass filter and the corresponding mFBP reconstruction is approximately
back-projection. 

We point out that there is other work on deriving filters for FBP
 applied to tomosynthesis \cite{Stevens2000,Kunze2007,Nielsen2012}.
Also, additional parameters such as a slice thickness filter or Hanning filter
width can be beneficial for controlling the DBT image quality. \cite{Orman2006,Mertelmeier2017}
Our particular design results from empirical subjective assessment on volumes reconstructed
with the Hologic geometry
and the aim of having a single-parameter image reconstruction algorithm.

\textbf{LSQI} - We also investigate whether orientation-dependence of fiber-like signal
conspicuity can be mitigated using iterative reconstruction algorithms.
Such algorithms employ a discrete model, and for the present work we employ a voxel expansion to arrive at
a discretized form of Eq. (\ref{Xraymodel})
\begin{equation*}
g = Xf,
\end{equation*}
where $f$ is a vector of $m$ voxel coefficients; $g$ is a vector of the $n$ line-integral measurements;
and $X$ is the system matrix generated by computing the projection of each of the voxel expansion elements.
The dimensions of the voxels used here measure 0.14$\times$0.14 mm$^2$ in-plane and 1.3 mm in depth. In using
$g$ and $f$ without parenthetical arguments, we are referring to the discrete digital data and voxel
coefficients, respectively. With parenthetical arguments, $g(\xi,u,v)$ and $f(\vec{r})$,
we are referring to the continuous data function and volume, respectively.

We consider two forms of least-squares optimization.
The first form of least-squares employs identity Tikhonov regularization (LSQI)
\begin{align}
	\label{eqn:LSQI}
	f_\text{recon}=\argmin_f \left \{ \|Xf - g\|^2 + \left( \lambda \|X\| \right)^2 \|f\|^2 \right \},
\end{align}
where $\lambda$ is a normalized regularization parameter and
$\|X\|$ is the maximum singular value of the system matrix $X$. For more implementation
details on LSQI for DBT, consult Ref. \onlinecite{Rose2017}. 

 In this particular formulation of LSQI, the parameter $\lambda$ is dimensionless and the same value
of $\lambda$ will incur the same level of regularization no matter what length units are used
in specifying $X$.
Some sense of the impact of $\lambda$ can be deduced by deriving
the small and large $\lambda$-value limits of the LSQI analytic solution. 

Setting the gradient of the objective function in Eq. (\ref{eqn:LSQI}) to zero, and solving for $f$ yields
a direct linear expression relating $f_\text{recon}$ to the data $g$
\begin{align*}
f_\text{recon} = \left(X^T X + (\lambda \|X\|)^2 I \right)^{-1} X^T g.
\end{align*}
This expression is useful for interpreting the effect of the regularization parameter $\lambda$.
The transpose of the projection matrix $X^T$ represents a form of discrete back-projection.
The small $\lambda$ limit tends to back-projection filtration (BPF)
\begin{align*}
	\lim_{\lambda \rightarrow 0} f_\text{recon} = (X^T X)^{-1} X^T g,
\end{align*}
where the discrete back-projection, $X^T$, is followed by filtration in the form $(X^T X)^{-1}$.
Scaled appropriately, the large $\lambda$ limit tends to back-projection \cite{Rose2017}
\begin{align*}
	\lim_{\lambda \rightarrow \infty} (\lambda \|X\|)^2 f_\text{recon} =  X^T g.
\end{align*}
In practice, $f_\text{recon}$ is obtained
by iterative solution of Eq. (\ref{eqn:LSQI}) using the conjugate gradients algorithm.


\textbf{LSQD} - The second form of least-squares presented is more commonly used in tomographic applications
than LSQI; namely, we consider least-squares with a quadratic roughness penalty \cite{Fessler1994} (LSQD)
\begin{align}
	\label{eqn:LSQD}
f_\text{recon} &=\argmin_f \left\{
\|Xf - g\|^2 + ( \lambda S )^2 \|\nabla f\|^2 \right\} \\
S &= \|X\|/\|\nabla\|  \notag
\end{align}
where $\nabla $ is a finite differencing matrix approximating the gradient operator. The 
scale factor $S=\|X\|/\|\nabla\|$ serves to make the regularization parameter $\lambda$ dimensionless.

The direct expression for the LSQD volume is
\begin{align*}
f_\text{recon} = \left(X^T X + (\lambda S)^2 \nabla^T \nabla \right)^{-1} X^T g.
\end{align*}
The small $\lambda$ limit of this expression is the same as that of LSQI, a form of
back-projection filtration.
The scaled large $\lambda$ limit, however, differs
\begin{align*}
	\lim_{\lambda \rightarrow \infty} (\lambda S)^2 f_\text{recon} = (\nabla^T \nabla)^{-1} X^T g.
\end{align*}
The expression $(\nabla^T \nabla)^{-1}$ is a form of blurring, and as a result the large
$\lambda$ limit of LSQD is blurred back-projection.


\textbf{Pyramidal image volume} - The Hologic system reconstructs images onto a pyramidal volume
 to correct for the effects of magnification due to the extremely limited angular range.
In terms of the continuous reconstructed volume, slices are displayed in the transformed coordinate system
$(x^\prime,y^\prime,z^\prime)$, where the transformed coordinates
are  related to the original Cartesian coordinates by
\begin{align}
	&x(x^\prime,y^\prime,z^\prime) = \frac{C -  z}{C} x^\prime, \label{sc1}\\
	&y(x^\prime,y^\prime,z^\prime) = \frac{C -  z}{C}  y^\prime,  \label{sc2}\\ 
	&z(x^\prime,y^\prime,z^\prime) =  z^\prime.\label{sc3}
\end{align}
The reconstructed volume is represented in this new coordinate system by the function
\begin{align*}
	f_\mathrm{pyr}(x^\prime,y^\prime,z^\prime) = f_\mathrm{recon}\left(\vec{r}(x^\prime,y^\prime,z^\prime)\right)
\end{align*}
Note that the transformed coordinates match the untransformed coordinates at the plane of the
detector, i.e. $z=0$. As $z$ increases, $x^\prime$ and $y^\prime$ adjust to counteract the magnification
of the corresponding $z$-plane.
Results for all four investigated algorithms are presented in the transformed coordinates.

\color{red}
\textbf{Background flattening} - Raw reconstructions of the breast exhibit a low-frequency variation when reconstructing
at larger regularization strengths. The effect is particularly strong near the skin line. This gray-level variation needs
to be removed in order to display low-contrast tissue structures over extended areas of the breast. In the
present work, we fit a low-order polynomial to the 2D slice images then divide the raw 2D slice image
by this fitted polynomial.
\color{black}

\subsection{Relative intensity metric for characterizing orientation-dependence of fiber-like signals}
\label{sec:metric}
\begin{figure}[t!]
	\centering
	\includegraphics[width=0.45\textwidth]{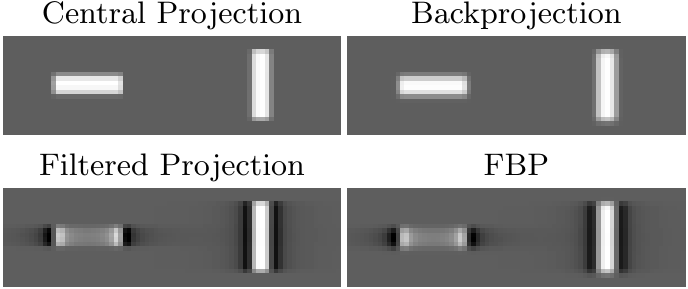}
	\caption{\label{fig:FBPOr} (Upper left)
The central projection, i.e. $\xi=0$, of 
DBT projection data generated for cylindrical rod signals
aligned parallel and perpendicular to the direction of X-ray source travel (DXST),
left-to-right in the image. The rod dimensions are 1.9 mm in length and
0.61 mm diameter. 
(Lower left) Ramp-filtered central projection. 
(Upper right)
Central slice of a reconstructed volume using back-projection. 
(Lower right)
Central slice of a reconstructed volume using filtered back-projection with the ramp filter. 
}
\end{figure}

\begin{figure*}[t!]
	\centering
	\includegraphics[width=0.72\textwidth]{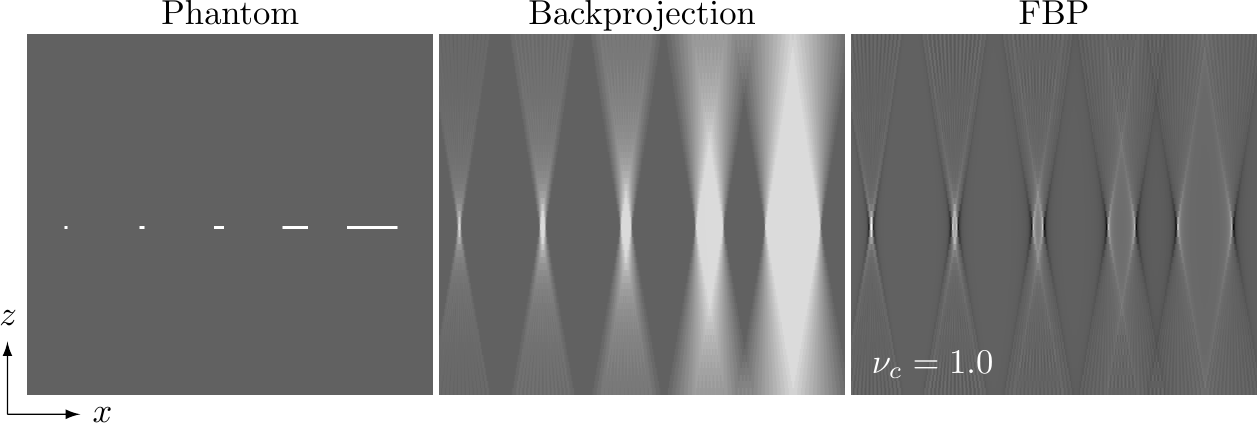}
\caption{\label{fig:roddepthbp}
Depth plane slice images of rods (left) parallel to the
direction of X-ray source travel (DXST): (middle) reconstruction by back-projection, and
(right) reconstruction by FBP. The rod lengths are 0.5, 1.0, 2.0, 5.0, and 10.0mm.
}
\end{figure*}

In order to provide background on fiber-like signal imaging in DBT
and to motivate the relative intensity metric, we show tomosynthesis images of a rod
aligned parallel and perpendicular to the direction of X-ray source travel (DXST) in
Fig. \ref{fig:FBPOr}. Shown are images for FBP using the ramp filtering
\begin{equation*}
H_\text{ramp}(\nu) =  |\nu|,
\end{equation*}
and no filtering
\begin{equation*}
H_\text{BP}(\nu) =  1,
\end{equation*}
which corresponds to back-projection (BP).
Because of the extremely limited angular range of the scan trajectory, the reconstructed slice images
resemble closely the central filtered projection prior to back-projection.

In the FBP image of Fig. \ref{fig:FBPOr}, the ramp filter causes two classic features of
objects reconstructed in a tomosynthesis setting\cite{Reiser2014,Sechopoulos2013}:
the ramp filter amplifies high frequency content in the $x$-direction causing the edges perpendicular to the DXST
to enhance, and it attenuates low frequency content in the $x$-direction causing gray-level contrast to drop toward
the middle of the signal. \color{black}
The impact of these FBP features differ greatly between
the images of the two rods. For the rod parallel to the DXST, the edge-enhancement is minimal
because the short edges are perpendicular to the DXST; furthermore, the gray-level contrast
is low relative to the actual rod because the long-axis is parallel to the DXST. For the rod
perpendicular to the DXST, the situation is reversed and the rod enhances well. These basic
imaging properties are a direct result of the ramp filter, and this is seen in comparing
the DBT slice images with the ramp filtered central projection in Fig. \ref{fig:FBPOr}.

The BP image shown in Fig. \ref{fig:FBPOr} does not have either the edge-enhancement nor
the drop in contrast in the signal interior. The central slice of the BP image appears to depict both rods faithfully.
In comparing the FBP to BP images, the former clearly have a greater variation with in-plane
orientation with respect to the DXST. In terms of conspicuity, BP will likely
yield uniform results with respect to orientation. For FBP, conspicuity may
vary greatly with orientation.

While the in-focus-plane images of Fig. \ref{fig:FBPOr} would seem to favor BP,
there is, however, a penalty for the absence of filtering; namely the depth-blur
of BP image reconstruction is significantly greater than that of FBP. The depth blur
of reconstructed rod signals of different lengths is shown in Fig. \ref{fig:roddepthbp} in a non-standard
viewing plane. The images are shown for planes aligned along the DXST and perpendicular to the detector.
The BP image strongly displays the classic size-dependent depth blur of DBT;
the BP depth blur has a central diamond shaped region of nearly uniform gray values, and outside
of this region the intensity slowly dissipates. The geometry of the central diamond region is determined
by the $x$-extent of the rod and the scanning angular range, which for the present configuration is 15$^\circ$.
For FBP, the reconstructed rod images in the depth plane have quite different appearances
from those of BP. The ends of the rod are highlighted, but for the longer
rods the ``BP diamond'' is suppressed including the mid-line where the actual rod is
located. 

Results from the investigated image reconstruction algorithms can be interpreted as some combination of
the two extreme cases of FBP and BP. There is a trade-off for fiber-like signals.
For FBP-like algorithms, there is strong orientation-dependence of the conspicuity in the focus plane.
For BP-like algorithms, there is pronounced depth blur for fibers parallel to the DXST.
To quantify this trade-off we employ a relative intensity metric defined through noiseless
simulation of rod-signals.

\begin{figure}[b!]
	\centering
	\includegraphics[width=0.45\textwidth]{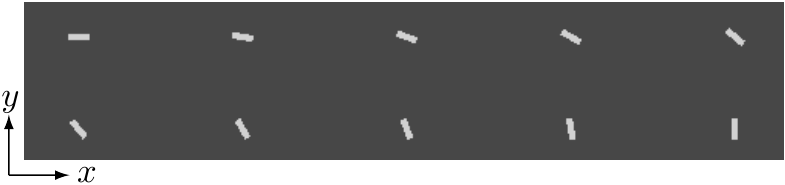}
	\caption{\label{fig:phant} Slice of simulation phantom employed in characterization studies of FBP,
mFBP, LSQI, and LSQD. The rod signals measure 1.9 mm in length and 0.61 mm in diameter. The rods are embedded
in a rectangular prism background measuring 11 cm $\times$ 4 cm $\times$ 4 cm. 
The phantom background attenuation
is chosen to be 0.629 cm$^{-1}$, modeling the average attenuation of fatty and fibroglandular tissue for 20 keV X-rays.
The rod attenuation is 0.802 cm$^{-1}$, modeling fibroglandular tissue for X-rays at the same energy.
The dimensions and contrast of the rod
are chosen to be typical of a segment of fiber-like structures in the breast. Multiple rods are embedded in the 
phantom at various in-plane orientations in order to obtain a visual assessment of the rod conspicuity as a function
of orientation angle.}
\end{figure}

\begin{figure*}[t!]
	\centering
	\includegraphics[width=0.7\textwidth]{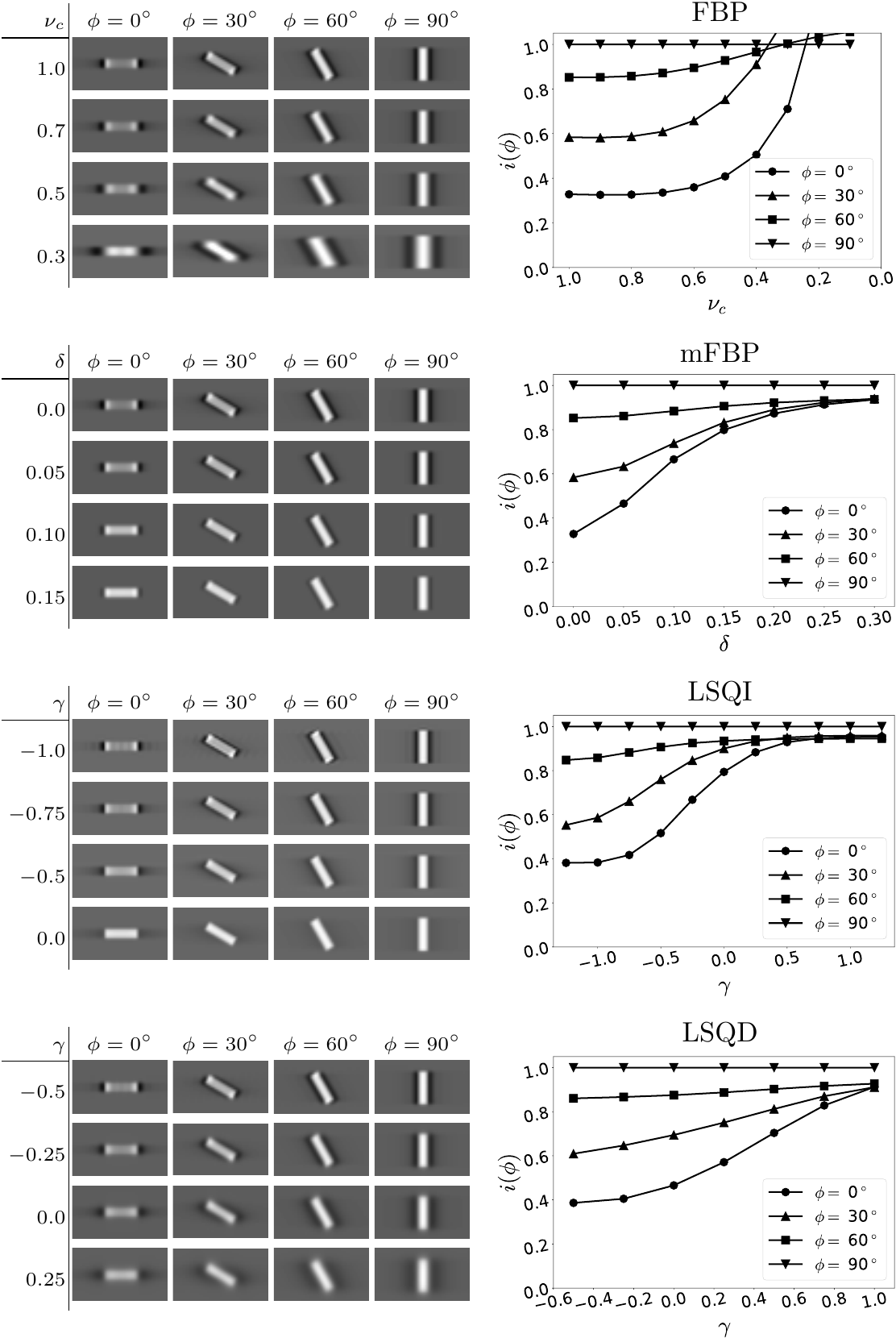}
	\caption{\label{fig:NoiselessSim}
	Left column: Region of interest (ROI) from the central slice of the reconstructed volume from noiseless 
	data of the rods-only phantom. Images of only 4 of the 10 rods are shown corresponding 
	to reconstruction from FBP, mFBP, LSQI, and LSQD at various regularization strengths, 
	where the regularization parameter setting for LSQI and LSQD is shown in terms of 
	$\gamma = \log_{10} \lambda$. Grayscale windows are different for each regularization 
	strength but constant for a single algorithm at one regularization strength. The maximum 
	value of the display window was chosen to be the maximum intensity of the $90^\circ$ 
	signal while the minimum value was chosen so a pixel value of 0 appears at approximately 
	the same luminance in each reconstruction. Right column: Plots of orientation-dependent 
	relative intensity for each algorithm and rod signal as a function of regularization strength. 
	Notation is simplified by defining ${i(\phi)= i(z=0,\phi,L=\SI{1.9}{\milli\metre})}$}
\end{figure*}

\subsubsection*{Orientation-dependent relative intensity}
We define a relative intensity metric to characterize reconstructed rod signal orientation-dependence
and depth blur.
Different reconstruction algorithms can yield large variation in gray scale values
yet have similar conspicuity of vessels, ducts, masses, and microcalcifications.
The variations in reconstructed gray scale also make comparison with the test object non-trivial \cite{Rose2017}.
We characterize rod signal fidelity with a metric derived from the reconstruction of a rod test
phantom shown in Fig. \ref{fig:phant}.
The background
tissue is removed and only the rods are used in generating noiseless projection data.

\begin{figure*}[t!]
	\centering
	\includegraphics[width=0.99\textwidth]{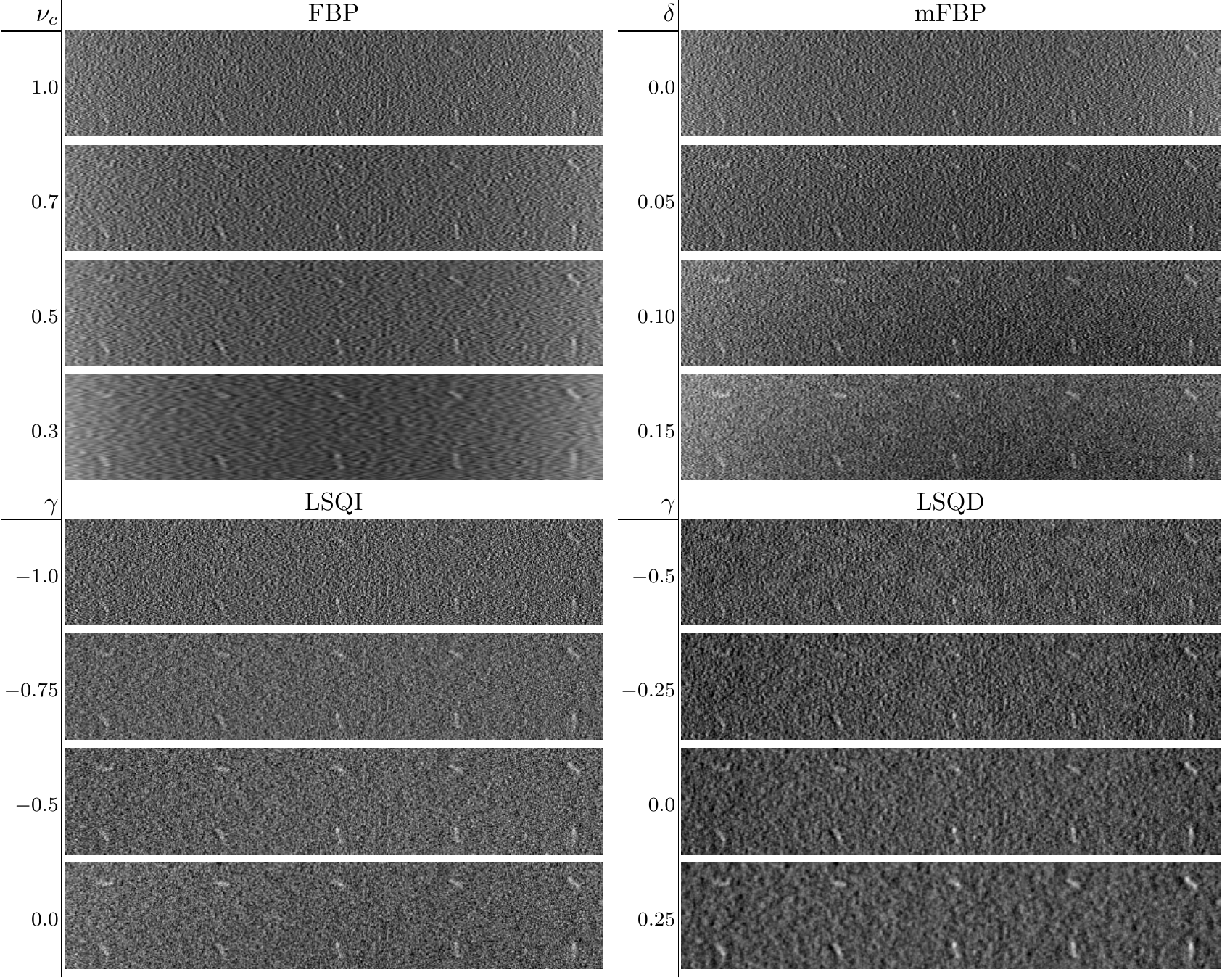}
	\caption{\label{fig:NoisySim}
Slice images of rods-plus-background phantom reconstructed from noisy data using each of the four considered
image reconstruction algorithms with four different settings of their respective regularization parameter. 
For each of the algorithms, the image regularization increases from top to bottom.
For LSQI and LSQD, the regularization strengths are shown in terms of $\gamma = \log_{10} \lambda$.
Grayscale windows were chosen subjectively for each reconstruction algorithm and regularization strength.
}
\end{figure*}

The relative intensity metric involves line-integration of the reconstructed rod volume,
$f^\text{rod}_\text{recon}$, along a line segment parallel
with the axis of the rod object
\begin{equation}
\label{rodIntensity}
I(z,\phi_{AB},L_{AB}) = \int_0^{L_{AB}} f^\text{rod}_\text{recon}(\vec{p}_{A}  + t \hat{\phi}_{AB}+ (z+z_A) \hat{e}_z ) dt,
\end{equation}
where $A$ and $B$ label the endpoints of the rod central axis;
$\vec{p}_{A}=(x_A,y_A,z_A)$ is the location of $A$;
$\phi_{AB}$ is the in-plane deflection of the rod axis from the DXST; 
the unit vector $\hat{\phi}_{AB}$ points in the direction of $B$ from $A$;
$L_{AB}$ is the rod axis length; and
$z$ is the depth separation from the rod axis. Note that $z=0$ results in integration along the rod axis.
This rod intensity measure applies to any rod signal, not just those shown in Fig. \ref{fig:phant}.
The only restriction imposed is that the rod axis must lie within a slice; i.e. $z_A=z_B$.

For the relative intensity metric, the line integration is normalized to that of
of $\phi=90^\circ$ and $z=0$
\begin{equation}
\label{rodRelIntensity}
i(z,\phi,L) = I(z,\phi,L)/I(z=0,\phi=90^\circ,L).
\end{equation}
This definition is motivated by the observation from Fig. \ref{fig:FBPOr}, that when orientation dependence is observed, the gray-level contrast is highest for the $\phi = 90^\circ$
rod. The relative intensity $i(z=0,\phi,L)$ for the rods in the phantom is 1.0, because all the phantom
rods are of the same length and gray level. Thus, any deviation from 1.0
directly reflects orientation-dependent intensity induced by the image reconstruction algorithm.
Ideally, $i(z,\phi,L)$ should be independent of $\phi$, 1.0 for $|z|<R$, and 0.0 for $|z|>R $, where $R$ is the rod radius.
Depth blur causes $i(z,\phi,L)$ to be larger than 0.0 for $|z|>R $.
The relative intensity metric tracks orientation dependence and persistence to out-of-focus planes
for reconstructed rod objects.

\section{Results}
\label{sec:results} 

\color{red}
The studies in Sec. \ref{sec:noiseless} and \ref{sec:noisysim} employ simulated data and
the relative intensity metric to illustrate how the orientation-dependence of fiber intensity depends on each algorithm's
single free parameter. Orientation-dependence of fiber conspicuity is presented with physical phantom 
data in Sec. \ref{sec:ACRorient}. The algorithm dependences
of the physical phantom images are shown in Sec. \ref{sec:ACRalgorithm}. Finally, the impact
of algorithm dependence on a clinical DBT case is presented in terms of fiber-like signal
conspicuity in Sec. \ref{sec:clinical}.
\color{black}


\subsection{Orientation dependence of rod-signal intensity}
\label{sec:noiseless}

Zoomed in reconstructions of a subset of the rods are shown in Fig. \ref{fig:NoiselessSim}
along with the orientation-dependent relative intensity metric. The shown images are in-plane
and the relative intensity metric is abbreviated as $i(\phi)$ with $z=0$ and $L=1.9$ mm.
$i(\phi)$ captures the trends in visual rod intensity relative to that of $\phi=90^\circ$ for
the various algorithms and regularization strengths. Overall for each of the image reconstruction
algorithms the deviation of $i(\phi)$ from 1.0 is greatest for low regularization strength and
small $\phi$. Increasing regularization reduces the orientation-dependent rod intensity as the
low $\phi$ rods show values of $i(\phi)$ approaching 1.0. 

The detailed behavior of each algorithm differs as each uses a different form of regularization.
The FBP Hanning filter regularizes by smoothing the image in the $x$-direction, and orientation-dependent
intensity is mitigated with the smoothing filter width, defined by $\nu_c$, comparable to the length of the rod signal.
While the intensity of $0^\circ$ and $90^\circ$ rods appears similar at large regularization (low $\nu_c$), the distortion of the rod has
strong $\phi$ dependence because of the directionality of the filter.
The mFBP filter parameter $\delta$ interpolates between ramp-FBP and BP, and the rod signal intensity and
shape become more uniform with increasing $\delta$.
The LSQI algorithm parameter $\lambda$ ranges between similar extremes as mFBP and accordingly
the resulting rod images are similar to mFBP. For LSQD, the low $\lambda$ limit is similar to that
of LSQI, but the large $\lambda$ limit involves smoothing of a different form than the other three algorithms,
which is reflected in the different appearance of the rod images.

\subsection{The orientation-dependence of small rod conspicuity}
\label{sec:noisysim}

The next set of simulation results illustrate how the intensity variation of fiber-like signals reconstructed
from noiseless data translate to variation in signal conspicuity in the presence of noise.
The simulated data are generated from projections with the complete
test phantom 
illustrated in Fig. \ref{fig:phant} that
contains a uniform background block and
10 rod signals oriented in angular increments of $10^\circ$ over a range of 
$0^\circ$ to $90^\circ$ with respect to the $x$-axis. 
Noise is modeled as an independent Poisson process in the transmission data with
a mean incident intensity of
$I_0 = 1.5 \times 10^4$ photons per ray.

The central slice from each reconstructed volume is shown in Fig. \ref{fig:NoisySim}.
Assessing the shown images for rod conspicuity yields an overview of in-plane orientation-dependence
and the impact of the various algorithms and forms of regularization. Starting with FBP in the upper left set
of images, there is a clear orientation-dependence of the rod conspicuity for $\nu_c=1.0$, the least regularized
case. The rod conspicuity increases with increasing angle to the $x$-axis, the DXST. Increasing regularization,
by decreasing $\nu_c$, visually seems to improve the conspicuity of the low-angle rods.

For each of the other three algorithms, mFBP, LSQI, and LSQD, at an overview level,
we observe similar trends to that of FBP for rod conspicuity. For the least regularized phantom image, there is
a strong orientation dependence with rod angle. Increasing regularization strength improves the conspicuity
of the rod-signals nearly parallel to the DXST.
These rod-conspicuity results are fairly consistent despite
the obvious differences in other image quality characteristics, such as noise texture or blur.


\subsection{Orientation-dependence of ACR-DM phantom fiber conspicuity}
\label{sec:ACRorient}

\begin{figure}[t!]
	\centering
	\includegraphics[width=0.48\textwidth]{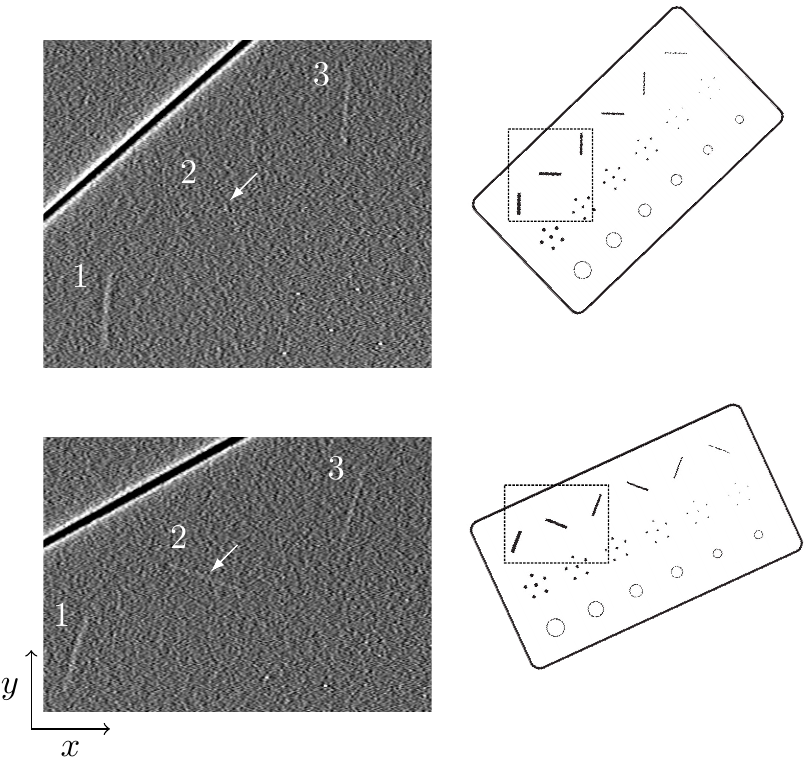}
	\caption{\label{fig:FBPEx} FBP reconstructions ($\nu_c = 1.0$) of the American College of Radiology
digital mammography accreditation phantom at two orientations.
In the top image the rods are oriented at approximately $5^\circ$ and
$95^\circ$ with respect to the $x$-axis. In the bottom image they are oriented at
approximately $15^\circ$ and $105^\circ$. Data for both orientations was acquired using
the same technique. Arrows in each ROI point to second thickest fiber, which exhibits
improved conspicuity in the $15^\circ$ orientation.}
\end{figure}

\begin{figure*}[t!]
	\centering
	\includegraphics[width=0.85\textwidth]{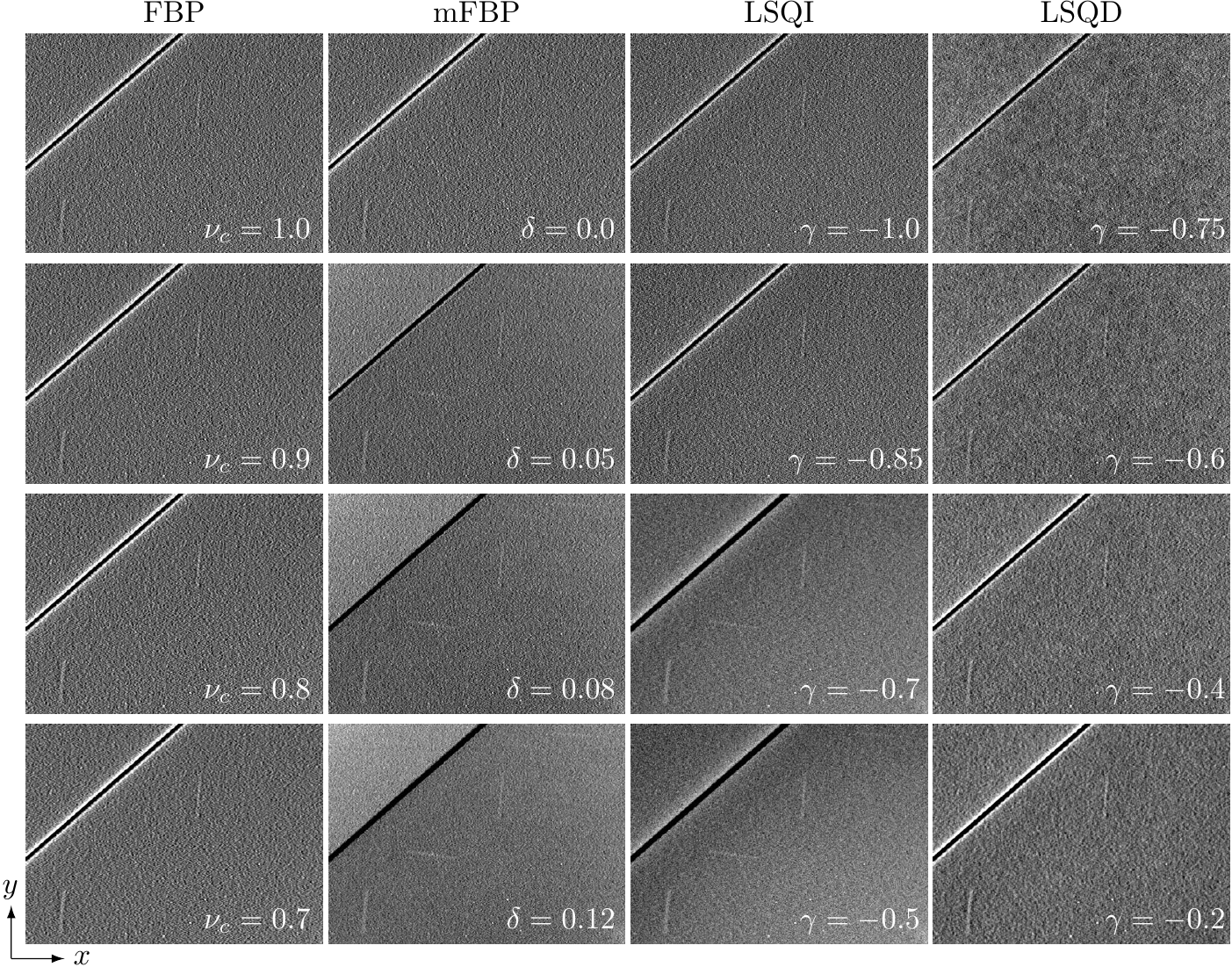}
\caption{\label{fig:ACRfocusplane} ACR-DM images for the in-focus plane. The phantom is tilted
by 40$^\circ$ from its standard orientation. The gray scale windows
are set subjectively for each panel.
}
\end{figure*}
To demonstrate the in-plane anisotropy of fiber-like signals, we show 
FBP reconstructions ($\nu_c = 1.0$) of the
ACR-DM phantom at two orientations in Fig.~\ref{fig:FBPEx}.
In 
Fig.~\ref{fig:FBPEx} the DXST is from left to right, and in the top row
the ACR-DM phantom is oriented at approximately a 40$^\circ$ angle
from the DXST. Orienting the phantom in this way causes the
odd-numbered fibers to line up nearly perpendicular to the DXST. The top-left
panel image shows the first three fibers, and the striking feature is that
the second fiber is invisible while fibers one and three are clearly seen.
Even more striking is that the fibers are numbered in order of decreasing
thickness, yet, fiber number three is seen and thicker
fiber number two is not. The bottom-row illustrates the imaging properties
of the same phantom when tilted 10$^\circ$ from the top-row orientation. 
This change in ACR-DM phantom orientation causes the second fiber to appear
in the bottom-left panel image.  The top-row image clearly displays how
fiber-like signals enhance when oriented perpendicular to the DXST
while those parallel do not; the bottom-row image hints at the strong dependence
of fiber-like signal conspicuity for orientations close to the DXST.


\subsection{Algorithm dependence of ACR phantom fiber conspicuity}
\label{sec:ACRalgorithm}

We expand upon the results shown in Sec. \ref{sec:ACRorient} for
the ACR-DM phantom tilted at $40^\circ$ so that the fiber objects line up at $5^\circ$ and
$95^\circ$ with respect to the DXST.
The ACR-DM phantom data is reconstructed by the four
considered algorithms at matched resolution with FBP.
The regularization equivalence among the four reconstruction algorithms is established
by matching the artifact spread function (ASF) \cite{Wu2004,Hu2008} of a simulated small
point-like object: a sphere of 28.0 $\mu$m diameter and attenuation coefficient of 9.29 cm$^{-1}$, modeling
a microcalcification. 

In Fig. \ref{fig:ACRfocusplane}, we show
ACR-DM images for the 40$^\circ$ orientation so that fiber signals at 5$^\circ$ and 95$^\circ$ are presented.
The 5$^\circ$ fiber is difficult to
see; it is visible in three of the mFBP images, two of the LSQI images, and
\color{red} none of the LSQD and FBP images \color{black}.
We note that there are differences in the noise texture among the images from the
various algorithms. In particular, that of LSQD is markedly different than that of the other three,
and the differences in noise texture can impact the fiber conspicuity.

\begin{figure*}[t!]
	\centering
	\includegraphics[width=0.85\textwidth]{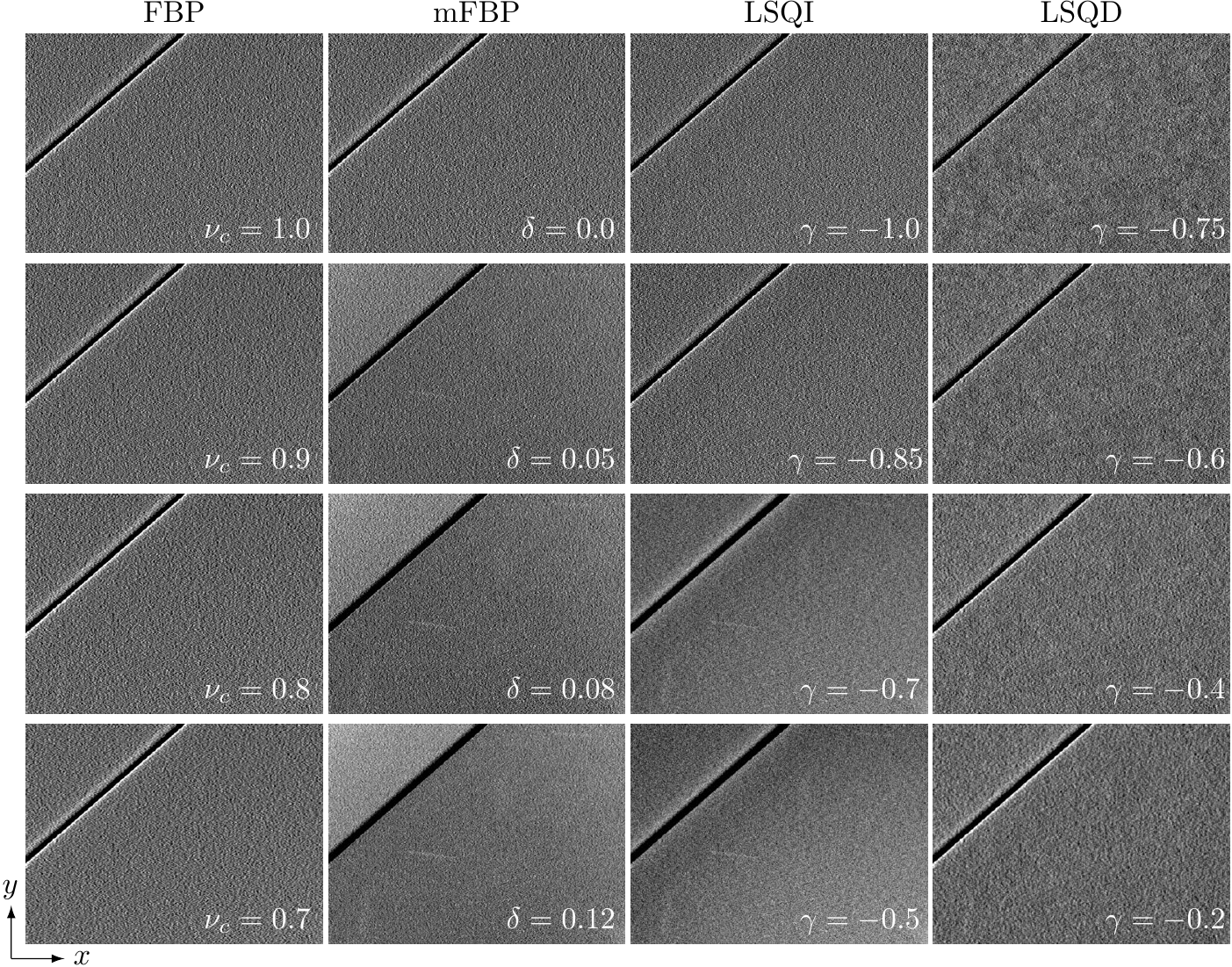}
\caption{\label{fig:ACRoutoffocusplane} Same as Fig. \ref{fig:ACRfocusplane} 
except ACR-DM images for a slice 7.7 mm below
the in-focus plane are shown. The gray scales are set to be exactly the same as
the corresponding image in Fig. \ref{fig:ACRfocusplane}.
}
\end{figure*}

Fig. \ref{fig:ACRoutoffocusplane} is identical with Fig. \ref{fig:ACRfocusplane} except
that the shown images are from a slice 7.7 mm below the focus plane for the 
ACR-DM fibers. 
The shown plane is well
below the fibers. The main point of viewing this out-of-focus plane is to observe the
extent to which fiber signals bleed through, yielding a sense of the object-dependent blur.
Most notable is the fact that the 5$^\circ$ fiber is still \color{red} visible \color{black} for the more
regularized images of mFBP and LSQI. Thus, it appears that the improved conspicuity of the 
$5^\circ$ fiber for the regularized mFBP and LSQI in-focus images comes at a price; namely,
this fiber persist strongly to out-of-focus planes.

\begin{figure*}[t!]
	\centering
	\includegraphics[width=0.8\textwidth]{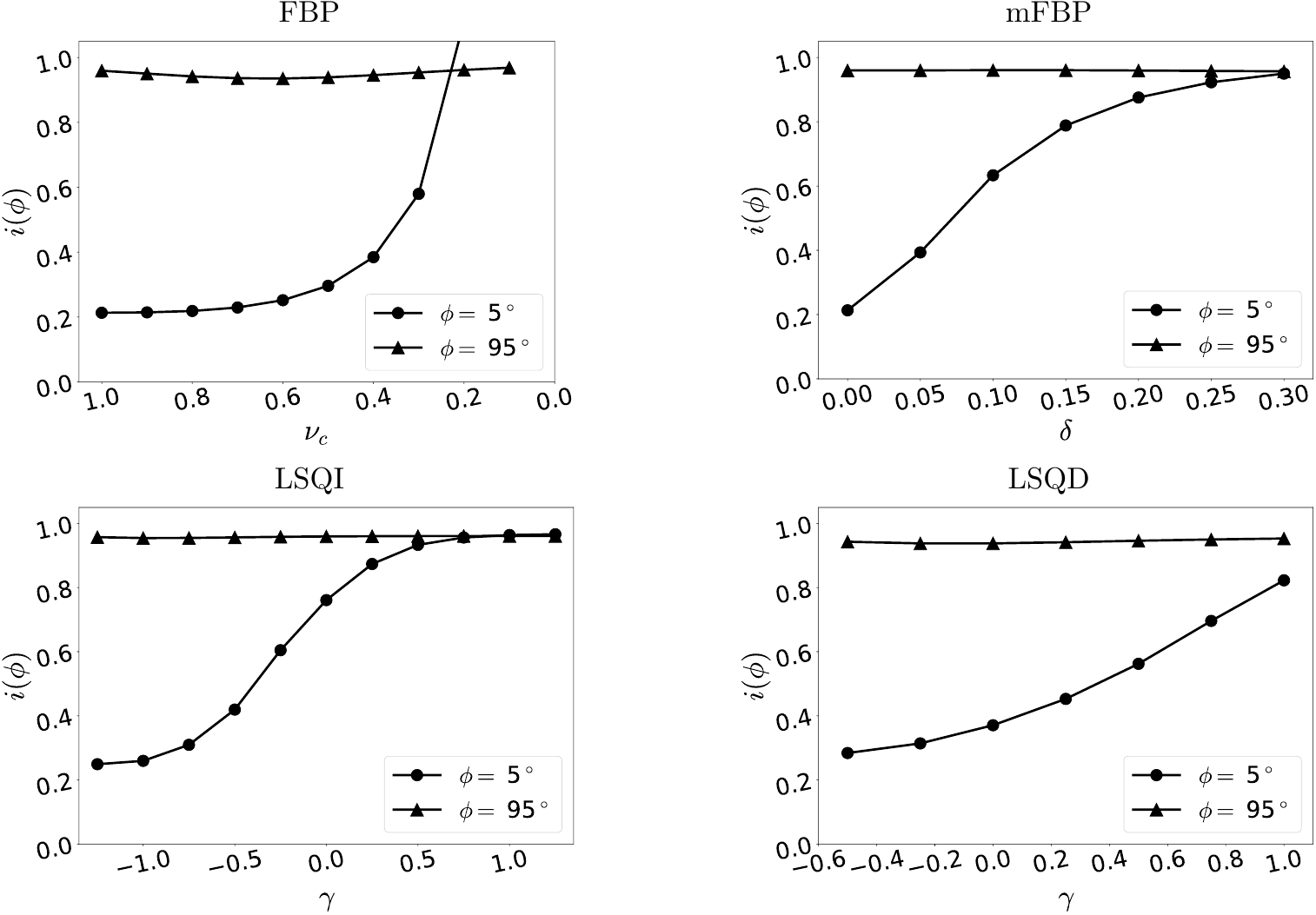}
\caption{\label{fig:rintensityFocus} Plots of relative intensity for ${z=0}$ and rods of length
${L=\SI{10.0}{\milli\metre}}$. The simplified notation ${i(\phi)=i(z=0,\phi,L=\SI{10.0}{\milli\metre})}$ is used. The rod diameter is 0.61 mm.
}
\end{figure*}
\begin{figure*}[t!]
	\centering
	\includegraphics[width=0.8\textwidth]{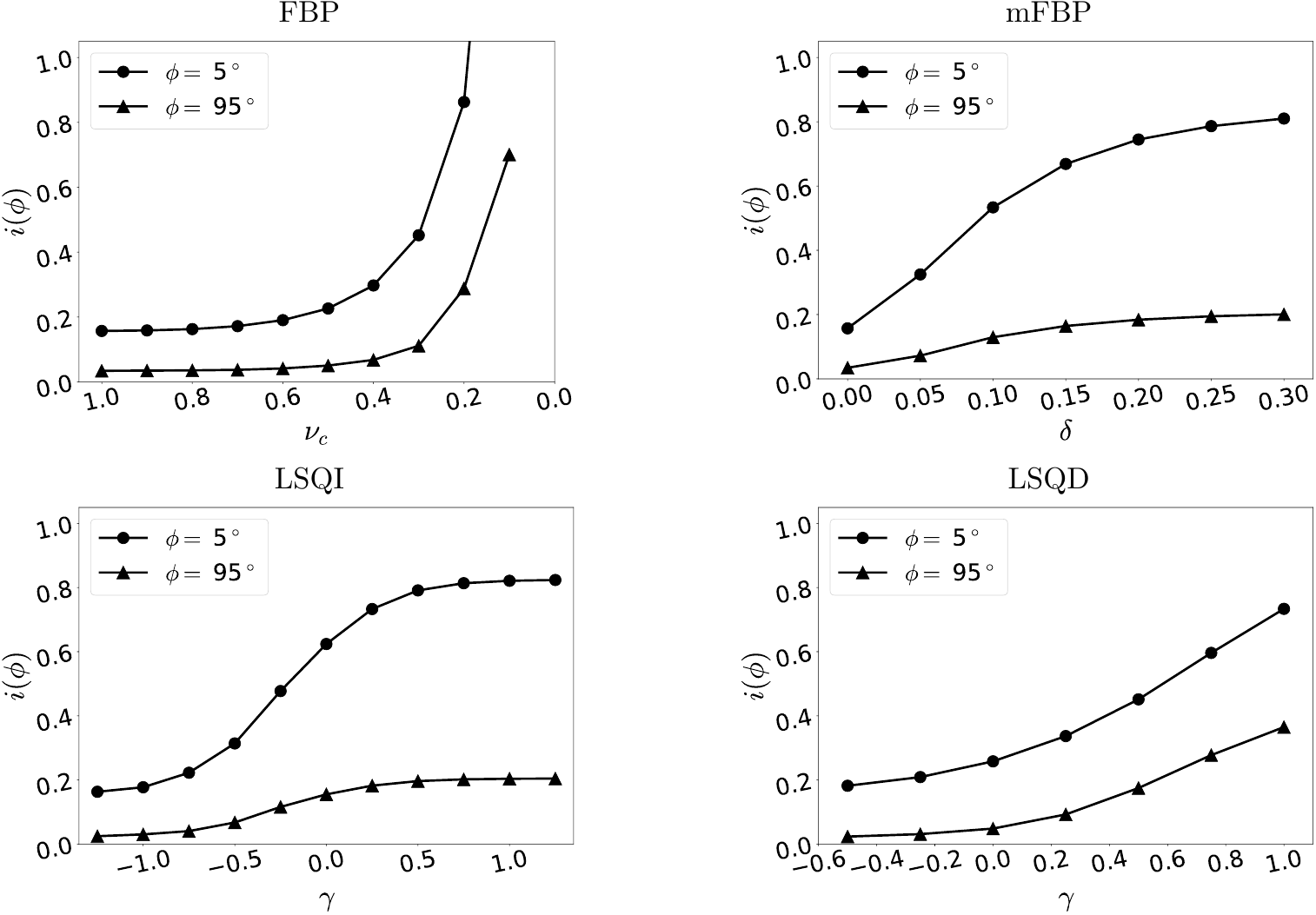}
\caption{\label{fig:rintensityOutoffocus} Same plots as those shown in Fig. \ref{fig:rintensityFocus}
except that the shown curves are for a plane $z= -7.7$ mm, i.e., 7.7mm below the in-focus plane.
The simplified notation ${i(\phi)=i(z=\SI{-7.7}{\milli\metre},\phi,L=\SI{10.0}{\milli\metre})}$ is used.
}
\end{figure*}
The relative intensity metric provides a means to interpret
the fiber conspicuity in the DBT scanner images of the tilted ACR-DM phantom.
For the present purpose, the metric is written $i(z,\phi)$ to allow for $z$-dependence
and the rod length is set to $L=10.0$mm to match the ACR phantom fiber length.
The in-focus plane
images in Fig. \ref{fig:ACRfocusplane} correspond to the relative intensity plots shown
in Fig. \ref{fig:rintensityFocus}.
Compared to the shorter  \SI{1.9}{\milli\metre} rods of Fig. \ref{fig:NoiselessSim} the dependences in
the relative intensity for the \SI{10.0}{\milli\metre} rods is more exaggerated. In particular, the rod signals
nearly parallel to the DXST have a much lower relative intensity than their shorter counterparts.
This agrees with the visual assessment of the fiber signals in Fig. \ref{fig:ACRfocusplane},
where the 5$^\circ$ fiber is difficult to see except when strong regularization is employed.
The images in Fig. \ref{fig:ACRoutoffocusplane}
shown for planes below the focus plane
correspond to the curves plotted in Fig. \ref{fig:rintensityOutoffocus}.
The relative intensity curves track the persistence of the small-angle ACR-DM fibers
from Fig. \ref{fig:ACRoutoffocusplane} seen for regularized mFBP and LSQI. The corresponding
relative intensity curves in Fig. \ref{fig:rintensityOutoffocus} increase with greater regularization
particularly for the low-angle rods reconstructed with mFBP and LSQI.
This result demonstrates
the trade-off between in-focus-plane orientation dependence of fiber conspicuity and depth blur of fibers
nearly parallel to the DXST.

\subsection{Fiber-like structure conspicuity algorithm dependence for a clinical DBT case}
\label{sec:clinical}

Up until the present section, we have focused on various aspects of the imaging of fiber-like structures embedded
in uniform backgrounds.
To investigate visualization of fiber-like signals in the presence of realistic background structure, 
we show an ROI from a clinical 
DBT case reconstructed by FBP, mFBP, LSQI, and LSQD. The chosen 
ROI has fibrous structures at different orientations so that
some of the discussed trends in rod signal conspicuity can be illustrated.

\begin{figure*}[t!]
	\centering
	\includegraphics[width=0.95\textwidth]{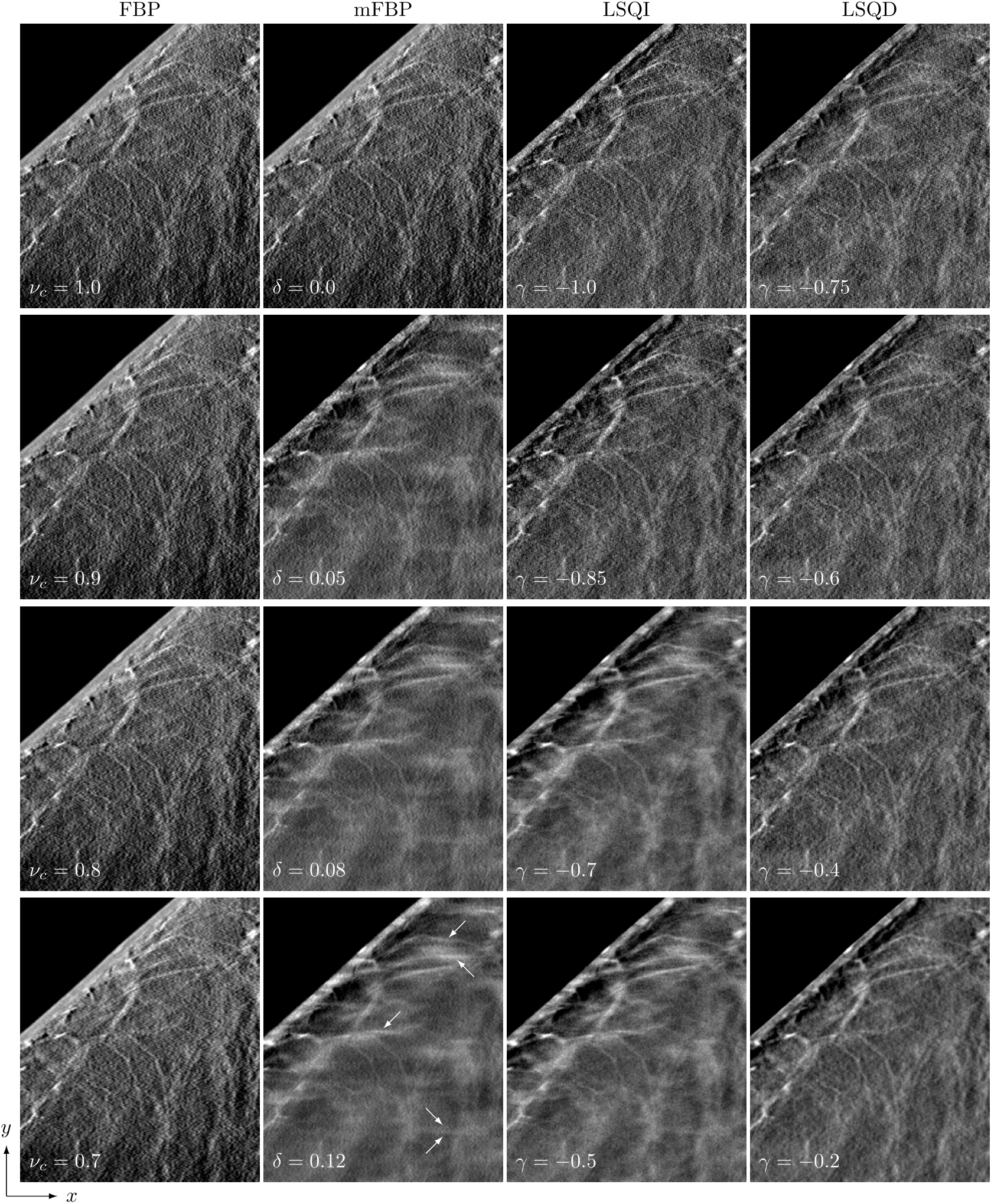}
\caption{\label{fig:clinical} An ROI from a DBT clinical case reconstructed by FBP, mFBP, LSQI, and LSQD
at various regularization strengths. For each row, the regularization strength is mapped to that of FBP by
matching depth resolution using the ASF of a simulated point-like object. The arrows in the mFBP,
$\delta=0.12$ panel indicate two features discussed in the text. The upper three arrows indicate a group of three
vessels that are nearly aligned with the DXST. The lower pair of arrows point to a horizontal shading artifact
visible in the regularized mFBP and LSQI images. The ROI has dimensions 
(\SI{3.8}{\centi\metre}$\times$\SI{4.5}{\centi\metre}).
}
\end{figure*}
In Fig. \ref{fig:clinical} an ROI from a clinical case is displayed in an arrangement similar to
the ACR-DM images shown in Fig. \ref{fig:ACRfocusplane}; the same ROI is shown for FBP, mFBP, LSQI, and LSQD
at the same regularization strengths used for the ACR-DM images. This particular ROI is chosen because of
the presence of multiple fiber-like structures at various in-plane orientations; and the conspicuity
of these structures can be compared among the algorithms and regularization strengths. In particular,
we highlight a group of three nearly horizontal vessels that change in conspicuity dramatically depending
on algorithm and regularization parameter setting. Their conspicuity varies in a manner consistent
with the presented rod-signal simulations and ACR-DM phantom results. 

The background texture in the patient images of Fig. \ref{fig:clinical} is seen to vary among the algorithms
and parameter settings, and the differences are more obvious than what is seen in the ACR-DM images
of Fig. \ref{fig:ACRfocusplane}. One aspect of the texture can be traced to imaging of fiber-like structures;
the mFBP images show horizontal wisp artifacts for $\delta \ge 0.05$ and to a slightly lesser degree the same
wispiness is seen in LSQI for $\gamma \ge -0.75$. These artifacts can be traced to vessels that are present
at slices of different depths. For example, the indicated horizontal wisp in the mFBP image for
$\delta = 0.12$ can be traced to a vessel oriented parallel to the DXST in a plane that is 1.3 cm below
the shown image. The horizontal wispiness is a direct result of the persistence of horizontal fiber-like
structures from other slices at different depths.



\section{Discussion and conclusion}
\label{sec:conclusion}

The clinical DBT images demonstrate that orientation-dependent conspicuity of fiber-like signals
can play a role in determining image reconstruction algorithm and regularization parameter settings.
To help guide the design of DBT image reconstruction algorithms, a relative intensity metric is proposed.
This metric quantifies the orientation-dependent intensity and depth blur of fiber-like objects.
It is a summary metric that is a descriptive property of noiseless reconstructions of rod signals,
and it is predictive of the gross behavior of conspicuity of fiber-like structures in the presence of noise.

The simulation trends agree well with visualization of the fibers in the ACR-DM phantom. The utility
of the simulation metrics is that they allow for quantitative comparison between algorithms and 
regularization settings; visualization of physical phantom images is subjective and affected by
the gray-level window settings and post-processing algorithms.
The trends in rod signal imaging revealed by the simulation
studies also help to explain features of imaging fiber-like structures in clinical DBT.
In particular, it is observed that fiber-like structures that run parallel to the direction
of X-ray source travel (DXST) can have their intensity suppressed relative to the similar in-plane
structures that are oriented obliquely to the DXST. Regularization that involves admixture of
back-projection can dramatically improve the conspicuity of fiber-like structures parallel to the DXST
at the expense of incurring depth blur of the same structures.

The rod signal relative intensity metric is motivated in this work by a need to explain variations
in fiber-like structure conspicuity. The natural extension of this idea is to develop model
observers for detection of the rod signals \cite{Diaz2015,Rose2017,RoseSPIE2018}. Even if such model observers
are developed, the present relative intensity metric can still provide useful information.
In particular, it can be directly applied to non-linear image reconstruction algorithm development
for DBT \cite{Sidky2009,Zheng2018,Garrett2018,Lu2010a,Ertas2013,Das2011,michielsen2013patchwork,Piccolomini2016}.

However, care must be taken to properly tailor the test phantom to
challenge the assumptions of the non-linear algorithm. For example, the uniform background
may lead to an optimistic assessment of total-variation (TV) based algorithms that exploit gradient sparsity.
In such a case it may be prudent to embed the rod signals in some form of structured background.

\color{red}
Finally, the DXST is seen to play a large role in imaging fiber-like structure; thus it may
be interesting to investigate rod signal conspicuity with different X-ray source trajectories.
\color{red}
For example, while strong orientation dependence of fiber-like signal conspicuity was observed in this 
work when reconstructing with low regularization settings, we would expect this effect to be 
less pronounced if X-ray projections were acquired over a larger angular span.
A $15^\circ$ source trajectory was considered in this work, but there exist clinical 
tomosynthesis systems employing angular ranges up to $50^\circ$. The variation
in depth blur and orientation dependent conspicuity as the regularization strength is 
increased would appear differently for such systems. Crosstalk between slices would, in general,
be expected to be less pronounced at all regularization strengths. We therefore emphasize that the particular 
results presented here concern the $15^\circ$ source trajectory. 

\color{black}

In conclusion,
we have investigated the impact of image reconstruction algorithm and associated parameter
settings on imaging of fiber-like structures. For the shown algorithms at a low regularization
setting, a strong orientation-dependence in the conspicuity of fiber-like signals is observed in 
simulation, physical phantom, and clinical data. In particular, fiber-like objects nearly parallel to
the DXST show a drop in conspicuity. Increasing regularization strength mitigates this orientation
dependence at the cost of increasing depth blur of these structures. 

\section{Disclosure of Conflicts of Interest}
The authors have received a research grant from Hologic, Inc. 

\section{Acknowledgements}
The authors would like to thank Drs. Chris Ruth,
Yiheng Zhang, and Chao Huang from Hologic for providing data and useful discussion. 
This work was supported in part by NIH R01 Grants Nos. CA182264, EB018102, EB026282 and 
NIH F31 Grant No. EB023076. The contents of this article are solely the responsibility of the authors 
and do not necessarily represent the official views of the National Institutes of Health.

\bibliography{library}



\end{document}